\documentclass[pdflatex, sn-mathphys-num]{sn-jnl}% Math and Physical Sciences Numbered Reference Style
\usepackage{siunitx}

\usepackage{upgreek}

\usepackage{graphicx}%
\usepackage{multirow}%
\usepackage{amsmath,amssymb,amsfonts}%
\usepackage{amsthm}%
\usepackage{mathrsfs}%
\usepackage[title]{appendix}%
\usepackage{xcolor}%
\usepackage{textcomp}%
\usepackage{manyfoot}%
\usepackage{booktabs}%
\usepackage{algorithm}%
\usepackage{algorithmicx}%
\usepackage{algpseudocode}%
\usepackage{listings}%

\usepackage{cleveref}
\begin{document}

\title{Differentiable Hybrid Modelling for Learning and Optimising Chemical Transport Processes from Experimental Data}


\author[1]{\fnm{Arthur} \sur{Jessop}}\email{arthur.jessop@manchester.ac.uk}

\author[1]{\fnm{Mohammed} \sur{Alsubeihi}}\email{mohammed.alsubeihi@manchester.ac.uk}

\author*[2,3]{\fnm{Ben} \sur{Moseley}}\email{b.moseley@imperial.ac.uk}

\author*[1]{\fnm{Ashwin} \sur{Kumar Rajagopalan}}\email{a.rajagopalan@manchester.ac.uk}

\affil[1]{\orgdiv{Department of Chemical Engineering}, \orgname{The University of Manchester}, \orgaddress{\city{Manchester}, \postcode{M13 9PL}, \country{United Kingdom}}}

\affil[2]{\orgdiv{Department of Earth Science and Engineering}, \orgname{Imperial College London}, \orgaddress{\city{South Kensington}, \postcode{SW7 2BX}, \country{United Kingdom}}}

\affil[3]{\orgdiv{I-X Centre for AI in Science}, \orgname{Imperial College London}, \orgaddress{\city{South Kensington}, \postcode{SW7 2BX}, \country{United Kingdom}}}


\abstract{
        Transport models underpin the design and optimisation of chemical processes, yet their constitutive laws are typically hand-picked, their experimental initial conditions are assumed exact, and their computational cost restricts optimisation to a handful of process variables.
        Here we introduce a differentiable hybrid modelling framework that embeds neural networks within a GPU-accelerated population balance equation solver to discover constitutive laws and assimilate initial conditions directly from experimental crystallisation data.
        Using thirteen real experiments with multi-dimensional size and shape measurements, we show that treating measured initial conditions as exact yields a physically inconsistent rate law despite improved predictive accuracy, whereas assimilating them recovers a physically consistent, non-linear law and reduces trajectory error sevenfold relative to conventional modelling.
        End-to-end differentiability further enables gradient-based optimisation of high-dimensional operating profiles, identifying a 300-variable temperature profile that reduces crystal aspect ratio by 25\% at constant mass, facilitating superior product attributes.
        Our framework augments, rather than replaces, mechanistic modelling, offering a general route to learning transport models from data and optimising processes with them.
}


\maketitle

\section{Introduction} \label{sec-intro}

Transport equations are the mathematical foundation for modelling dynamic systems across chemical engineering.
They describe how mass, energy, and momentum are transported within complex systems.
Applications range from predicting the rheology of consumer products \cite{edwards1991interfacial}, to quantifying adsorption processes for carbon capture \cite{rajagopalanAdsorbentScreeningMetrics2016}, analysing solid-form products in pharmaceutical and agrochemical processes \cite{ramkrishna2000population}, and guiding the design of fluidized beds \cite{Kunii1991Fluidization}, bioreactors \cite{mcduffie1991bioreactor}, and heat exchangers \cite{sekulic2023fundamentals}.
Process models built on these equations let engineers explore design spaces, predict, optimise and troubleshoot processes, and implement model predictive control \cite{aprileReflectingBarriersContinuous2025, kocevskaChangingParadigmIndustrial2024}.
Accurate and expressive models are therefore key to the digital and autonomous manufacturing goals of Industry 4.0 and 5.0 \cite{akundiStateIndustry50Analysis2022}.

To model real systems, constitutive laws must be fit within these transport equations to accurately capture system dynamics.
Traditionally these equations are hand-picked by experts and fitted via trial-and-error for every new problem \cite{botschiFeedbackControlSize2019}.
However, selecting a fixed functional form inherently limits the model to a physical interpretation that may not capture the true underlying phenomena.
A second assumption that is often made and rarely stated is that the initial conditions from which the laws are learned are exactly known.
In practice these are experimentally measured and therefore subject to noise or inherent measurement device limitations but often treated as ground truth.
Yet, for dynamical systems where one tracks a quantity of interest over time, solutions can be very sensitive to initial conditions \cite{lorenzDeterministicNonperiodicFlow1963}.
Together these assumptions can significantly bias model predictions and introduce systematic errors in real-world prediction and optimisation workflows, limiting experimental efficiency and slowing model development.

Even with a reliable model, using it for process optimisation presents challenges.
Typically, transport models are solved with numerical solvers that are computationally expensive \cite{ParallelHR2008} and output only forward solutions.
Optimisation workflows therefore use gradient-free or black-box algorithms, or finite-difference gradients, whose cost (solver calls) grows with the number of process variables.
This limits exploration of the high-dimensional design spaces of real processes, forcing reliance on simplified optimisation strategies or experimental trial and error, potentially leaving improvements undiscovered.

While these issues affect all transport equations and process models built upon them, this work focuses on population balance equations (PBEs) \cite{ramkrishna2000population, ramkrishnaPopulationBalanceModeling2014}.
These are number balance equations frequently applied to crystallisation, a unit operation common in the pharmaceutical, agrochemical, and fine chemicals industries to manufacture solid-form products \cite{ramkrishna2000population}.
Formulated as partial differential equations (PDEs), they track the number of particles of a given size and shape to model how the overall particle size and shape distribution (PSSD) evolves over time.
We use crystallisation as our demonstration system because the issues raised above are highly relevant to both academia and industry, alongside additional challenges specific to this particular problem.
Process kinetics are unknown and system-specific \cite{Perspectives2020, kocevskaChangingParadigmIndustrial2024}, meaning a new hand-picked model must be built for every system.
Most commercial monitoring tools that provide data for model building also reduce particle shapes down to a single dimension, where a multi-dimensional approach would be more descriptive and facilitate the recovery of multi-dimensional laws \cite{dealbuquerqueEffectNeedlelikeCrystal2016}.
Additionally, a recent comparative study has shown that popular measurement tools frequently disagree with one another and with ground-truth measurements of initial conditions and other state variables \cite{biriThereRightParticle2025}.
Finally, conventional crystallisation optimisation is often treated as a black-box problem, with optimisation limited to a small number of operating variables such as ramp rates or temperature cycles \cite{barhateDigitalDesignCrystallization2026, orosz2DPopulationBalance2025a}.

Recent work has begun to use machine learning (ML) to address these limitations individually, either through purely black-box methods \cite{barhateDigitalDesignCrystallization2026, raponiMultivariateOptimizationInverse2026, orosz2DPopulationBalance2025a, zhengMachineLearningModeling2022}, physics-informed ML approaches \cite{mukherjeePhysicsConstrainedMachineLearning2025, sharmaHybridScienceguidedMachine2022, pessinaTransferLearningDatadriven2026, wuPhysicsinformedMachineLearning2023, naiModelingBatchCrystallization2026} or hybrid approaches \cite{vonstoschHybridSemiparametricModelling2014, limaImprovedModelingCrystallization2023}.
However, existing approaches generally address only subsets of the challenges identified above and do not provide a single framework for learning unknown constitutive laws, assimilating uncertain initial conditions, forward simulation, and direct process optimisation \cite{picklesBridgingDataTheory2026}.
Moreover, many of these models are largely data-driven and can violate physical laws, provide limited physical insight, and remain specialised to the systems and quantities represented during training which is particularly problematic in highly regulated industries such as pharmaceuticals \cite{chewAIChemicalEngineering2026}.

\begin{figure}[htbp]
    	\includegraphics[width=\linewidth]{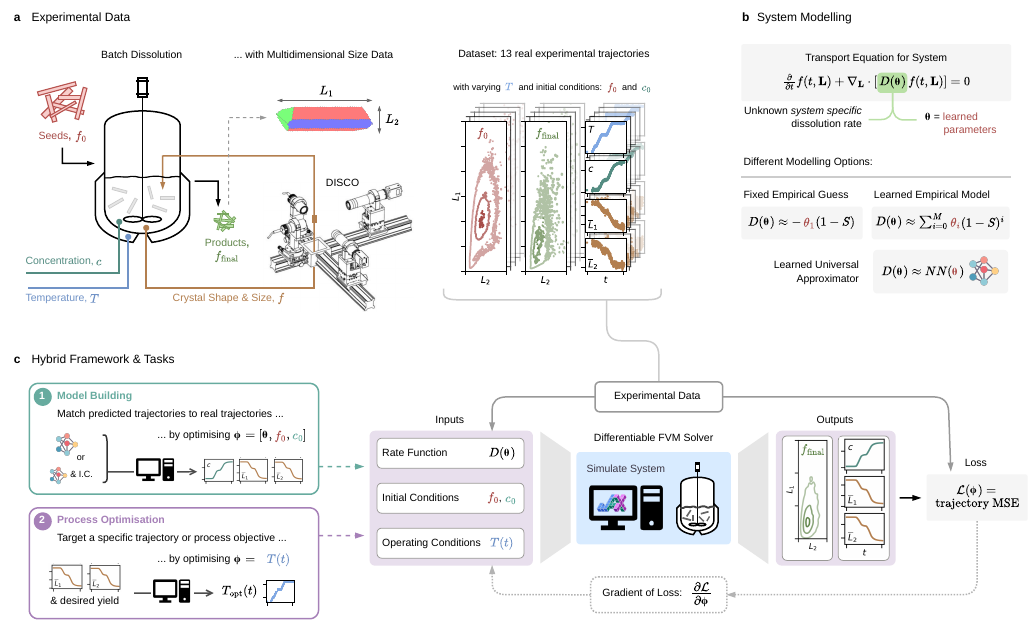}
		\caption{
        \textbf{A single, unified, differentiable hybrid framework for discovering transport laws, assimilating initial conditions, and optimising process trajectories.}
        \newline
        \textbf{(a)} Experimental system. Batch temperature-driven dissolution of an amino acid (\textsc{l}-glutamic acid) in water, monitored online using the DISCO \cite{rajagopalanComprehensiveShapeAnalysis2017} which characterises crystals in two characteristic lengths (length $L_1$, width $L_2$). The dataset contains 13 real experimental trajectories \cite{botschiFeedbackControlSize2019} with varying temperature profiles and initial conditions ($f_0$, $c_0$). Example seed ($f_0$) and product ($f_{\textrm{final}}$) distributions are shown alongside time series data.
        \textbf{(b)} System modelling. The simplified batch dissolution population balance equation is shown with the learned dissolution law ($D$) parametrised by $\boldsymbol{\uptheta}$ highlighted. Different options for formulating the law are shown and we demonstrate empirical and NN options in this work.
        \textbf{(c)} Hybrid framework and tasks. All inputs pass through the differentiable solver to give predicted outputs, and automatic differentiation is used to propagate gradients back from the trajectory mean squared error (MSE) loss function $\mathcal{L}(\boldsymbol{\upphi})$ to any subset of these inputs. This supports two tasks: model building, where predicted trajectories are matched to observed trajectories by optimising $\boldsymbol{\upphi} = [\boldsymbol{\uptheta}, f_0, c_0]$, and process optimisation, where a specific trajectory or process objective is targeted by optimising $\boldsymbol{\upphi} = T(t)$ or $[T(t), f_0, c_0]$.}
		\label{fig1}
\end{figure}

Here, we introduce a single, unified, differentiable, physics-based hybrid modelling framework (shown in \Cref{fig1}(c)) that learns constitutive laws and assimilates initial conditions directly from experimental data, rather than relying on empirical assumptions, and is capable of scalable, efficient gradient-based process optimisation.
The framework integrates a GPU-accelerated, JAX-based finite-volume PBE solver \cite{bradburyJAXComposableTransformations2018, alsubeihiModernEfficientDifferentiable2025} with learnable neural networks, which are trained to discover constitutive laws from experimental data while the initial conditions are jointly optimised alongside the network parameters.
The resulting hybrid model is end-to-end differentiable, meaning that we can leverage automatic differentiation and GPU acceleration to enable efficient gradient-based optimisation of model and process design parameters, with computational cost that scales effectively independently of the number of parameters being optimised.

We demonstrate the utility of the framework on thirteen real dissolution experiments \cite{botschiFeedbackControlSize2019} with two-dimensional crystal size and shape measurements.
The learned hybrid model identifies a non-linear rate law that improves predictive accuracy compared with a conventional hand-picked law \cite{botschiFeedbackControlSize2019}.
The jointly inferred initial conditions provide a better fit to the experimental data than the experimental initial conditions, indicating that the framework can assimilate partially uncertain experimental conditions.
Finally, we apply the learned model to process optimisation, where its end-to-end differentiability enables efficient and scalable optimisation over a large number of process variables.
For an example temperature cycling process, we identify a high-dimensional operating temperature profile that reduces product aspect ratio by 25\% at constant mass, enabling targeted control of crystal morphology that would be computationally prohibitive with conventional workflows.

\section{Results} \label{sec-results}

To demonstrate our hybrid modelling framework, we study the dissolution of the amino acid (\textsc{l}-glutamic acid) in water, where particles are characterised in multiple dimensions, their length and width ($L_1$, $L_2$), using the DISCO \cite{botschiFeedbackControlSize2019, rajagopalanComprehensiveShapeAnalysis2017}.
The driving force is the supersaturation $S$ [-], which is a measure of how far the system is from equilibrium, and the operating condition is the temperature $T$ [$^\circ\text{C}$].
If $S=1$, the system is at equilibrium, if $S<1$ then crystals dissolve (shrink), and if $S>1$ then crystals grow.

The system suits our framework: the experiments follow a workflow representative of industrial process development and combine condition-dependent kinetics with a multi-dimensional particle population.
Applying this framework to multi-dimensional particle populations is rarely attempted, in part because the required experimental data and computationally efficient solvers remain scarce.
Learning such kinetics from data is fundamental to developing predictive models for new processes.

\subsection{Data-Driven Rate Law Discovery and Initial Condition Assimilation} \label{ssec-predictive_performance} \label{ssec-model_building}

First, we use our hybrid framework to discover rate laws and assimilate initial conditions (IC) for \textsc{l}-glutamic acid dissolution using data from 13 experiments \cite{botschiFeedbackControlSize2019}, reported previously by the senior author.
Conventionally, candidate rate laws are selected manually and fitted assuming the measured initial conditions are exact.
Both of these assumptions can bias the fitted kinetics.

We benchmark against M1 (\Cref{eq:M1_coefficients}), the linear empirical rate law selected by B\"otschi et al. \cite{botschiFeedbackControlSize2019} from four candidate equations (M1 - M4), and evaluated in our solver using their published coefficients.
Testing four candidates for each characteristic length gives sixteen combinations, and this grows rapidly as further phenomena are added.
The model is also limited to the candidate forms tested, so a law outside of these cannot be described.
Instead, a neural network is embedded directly in the differentiable solver replacing the hand-picked rate law with a flexible function for $D(S,T)$ to make a \textit{Hybrid Model} (\Cref{fig1}(b)).
The network learns both $D_1$ and $D_2$ in a single training effort and can explore a broad functional space \cite{hornikMultilayerFeedforwardNetworks1989} without the need to manually supply or fix the rate to a particular empirical form.
This allows us to resolve kinetics in two dimensions within the time constraints of real-world process development, a search that is rarely attempted due to the lack of sufficiently fast solvers for multi-dimensional PSSDs that can support ML methods.

When trained traditionally with a train/validation data split and the measured initial conditions held fixed (\textit{Case A}, \Cref{ssec-training_methods}), the hybrid model lowers the validation loss (0.812) versus the fixed empirical baseline (1.264), where validation loss is a mean squared error (MSE, \Cref{eq:loss}).
However, across 20 randomly initialised models the learned rate is physically inconsistent (\Cref{exFig2}(a), blue line and band): it is non-zero as $S\to1$ and shows an inverted temperature dependence, dissolution rate decreasing as temperature increases.
The hybrid model can therefore improve predictive performance but encode the wrong physics, which is why predictive accuracy alone is not always a sufficient test.
As no structural element of the network enforces the rate to approach zero at equilibrium (\Cref{ssec-nn_methods}), equilibrium behaviour must be learned from the data, so we next examine whether the inconsistency stems from the model itself or the experimental data.

\begin{figure}[h]
    \centering
    \includegraphics[width=\linewidth]{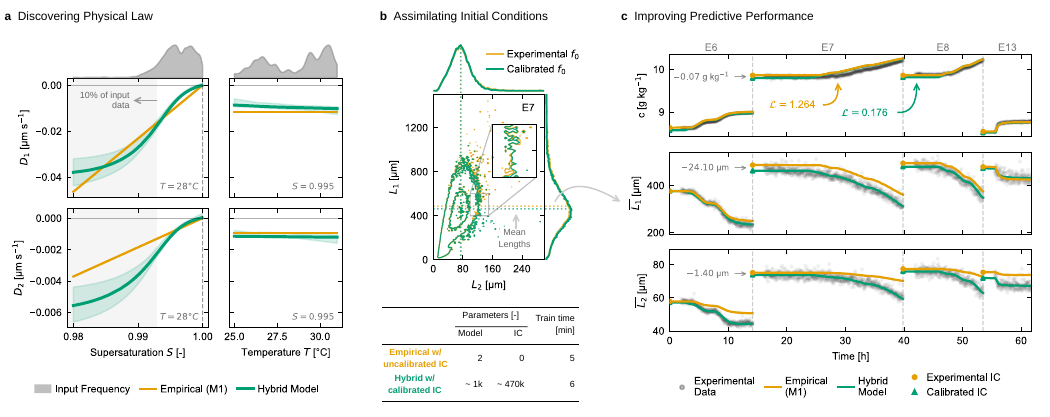}
    \caption{
            \textbf{Discovering a physically consistent, non-linear dissolution law from joint trajectory modelling and initial condition assimilation (calibration).}
            \textbf{(a)} Assimilating ICs recovers physically consistent kinetics. Learned dissolution rates ($D_1$, $D_2$) as functions of supersaturation (at fixed $T = 28\,^{\circ}\mathrm{C}$) and temperature (at fixed $S = 0.995$) for the calibrated-IC hybrid model (green) and the empirical baseline M1 (yellow). Solid line: ensemble mean; band: min/max over 20 random initialisations. Input frequency density is shown above each axis; the shaded region marks sparsely sampled $S$. The hybrid recovers a non-linear $S$ dependence that vanishes at equilibrium ($S = 1$) and negligible temperature dependence in line with M1, whereas the fixed-IC model learns non-physical kinetics (\Cref{exFig2}(a)).
            \textbf{(b)} Small corrections to the ICs. Measured (yellow) and calibrated (green) seed distributions for E7, the validation experiment with the largest shift, with marginal $L_1$ and $L_2$ distributions and their means (dashed). Assimilation only adjusts the shape, as the calibrated distribution is renormalised to the measured seed mass. Table: the parameter count increases by five orders of magnitude with little change in the time to optimise for 500 epochs, which is what makes IC assimilation feasible with AD.
            \textbf{(c)} Correcting the ICs drives the improved fit. Predicted time series of $c$, $\overline{L}_{1}$ and $\overline{L}_{2}$ across the validation experiments for M1 with measured ICs (yellow) and the hybrid model with calibrated ICs (green). Calibrated ICs shift only marginally, yet M1 with calibrated ICs alone reaches a trajectory loss of $0.195$ (\Cref{exFig2}), so most of the improvement comes from the ICs rather than the rate law.
            }
    \label{fig2}
\end{figure}

In multiple experiments, the ICs, particularly those of $\overline{L}_1$ and $\overline{L}_2$, do not match the point where the experimental data would back-extrapolate to (\Cref{fig2}(c), yellow circles vs grey dots).
This suggests measurement error in the ICs, and to reduce the training loss it may be preferable for the network to learn a rate that works to correct this error rather than describe the underlying law.
To address this, we leverage the differentiability of our solver to learn more accurate ICs, including them as uncertain free variables that are assimilated in parallel to training the hybrid model (\textit{Case B}), testing whether the mismatch influences the learned kinetics.
As each IC must be calibrated individually, \textit{Case B} trains on all experiments except the two held out for optimisation (\Cref{ssec-experimental_data_methods}).
We still report the trajectory loss on the original validation experiments for comparison with the empirical baseline, but as it is no longer a held-out loss, this is a comparison of modelling workflows rather than of rate laws.

This optimisation involves $\approx 4.7\times10^5$ parameters, dominated by the 11 PSSDs ($360\times120$ each).
When exploiting the automatic differentiation (AD) capabilities of JAX, 500 optimisation epochs can take between 5 - 6 minutes when training both the two-parameter M1 model (refit for timing only) and the full hybrid model with IC assimilation (\Cref{fig2}(b)).
For the benchmark MATLAB model of B\"otschi et al. \cite{botschiFeedbackControlSize2019}, which uses numerical gradients, fitting the two-parameter model can take upwards of 15 minutes, over twice as long as the full hybrid model despite optimising five orders of magnitude fewer parameters (\Cref{exFig4}).
AD makes gradient cost effectively independent of parameter count, and allows us to treat the ICs, and not only the rate law, as uncertain and assimilate them from the data.

Through this joint optimisation the recovered kinetics become physically consistent, \Cref{fig2}(a).
The $S$ dependence of the hybrid model now correctly approaches zero at $S=1$, and the temperature artefact is also removed, with the models exhibiting negligible temperature dependence consistent with M1.
The trajectory loss \Cref{eq:loss} on the original validation experiments is reduced to \num{0.176}, a sevenfold improvement, while assimilating the ICs with the published M1 law alone gives \num{0.195}, so most of this improvement arises from correcting the ICs rather than from the rate law itself.
During the joint optimisation the calibrated ICs only shift marginally from their measured values, as the data assimilation only adjusts the shape, not the mass corresponding to the distribution, which is renormalised during simulation (\Cref{ssec-training_methods}).
For E7, the validation experiment with the largest shift (\Cref{fig2}(b)), the experimental and calibrated distributions are quite similar, with the mean length and width shifting only \qty{24.1}{\micro\meter} (4.6\%) and \qty{1.4}{\micro\meter} (2.0\%), and a smaller shift still in concentration.
These shifts are within the noise in the rest of the trajectories, yet account for most of the improvement in predictive performance.
Our ability to assimilate ICs is therefore significant, as this effect cannot be seen with the conventional approach, where the measured ICs are assumed to be exact and assimilating them is computationally intractable (\Cref{exFig4}).

The learned function is of the same order of magnitude as the empirical baseline whilst revealing a non-linear response in $S$ that the linear M1 cannot describe (\Cref{fig2}(a)).
As expected, the ensemble spread of 20 models is narrow where $S$ is sampled densely and wider where it is sampled less densely.
In the densely sampled region the learned function differs in shape to the empirical equation, suggesting that the non-linearity is present in the data rather than imposed by the network.

\subsection{Scalable End-to-End Process Optimisation}  \label{ssec-process_optimisation}

Next, we use the trained framework for process optimisation, considering two optimisation tasks. 
First, we recover the temperature profile of a held-out experiment from its size trajectories alone, using the hybrid model introduced in \Cref{ssec-model_building}.
This is a common design task, and a real experiment provides a known answer to check against.
Second, we design a new temperature profile to reduce product aspect ratio (\Cref{sssec-aspect-ratio-optimisation}), using the published size-dependent growth and dissolution kinetics \cite{botschiFeedbackControlSize2019, ochsenbeinGrowthRateEstimation2014} (both growth and dissolution are required to manipulate aspect ratio and the hybrid model we train is purely size-independent dissolution).
This reflects a central industrial goal: achieving defined product specifications that satisfy regulatory requirements and facilitate downstream processing and the manufacturability of solid-form products.

In both cases, the operating condition is parametrised by 300 temperatures - significantly more degrees of freedom than conventional finite-difference (\Cref{exFig4}) and non-differentiable \cite{barhateDigitalDesignCrystallization2026, botschiFeedbackControlSize2019} workflows can practically accommodate.
A forward-difference gradient over 300 temperatures requires 301 forward solves, corresponding to $\approx$4 min per iteration at the reference solver's forward-simulation time (\Cref{exFig4}), and more than a day for a full optimisation.
With AD, as described in \Cref{ssec-model_building}, the gradient cost is effectively independent of the number of temperatures - with our framework, calculating the gradient with respect to the entire temperature profile takes approximately \qty{0.75}{\second}.
This removes the need to enforce a low-dimensional representation of the profile and allows the optimiser to identify profiles at the temperature controller resolution.

\subsubsection{Process Trajectory Optimisation} \label{sssec-trajectory-optimisation}

Often the operating condition, $T(t)$, is an imposed quantity manually defined to achieve a desired process trajectory - mirroring the philosophy of empirical equation selection, where the definition is driven by testing, expertise and process knowledge.
Here we directly optimise $T(t)$ toward a desired trajectory, using the hybrid dissolution model built in \Cref{ssec-model_building}.
We target crystal length and width trajectories ($\overline{L}_1$ and $\overline{L}_2$) which are measured by the DISCO \cite{rajagopalanComprehensiveShapeAnalysis2017} and directly related to product specification; concentration is not a  specification target here.

We first validate the task on synthetic data, separating the optimiser's ability to recover $T(t)$ from any mismatch due to model or IC uncertainty before testing on experimental data.
Target trajectories are generated by defining operating conditions (a stepped ramp and a sigmoid), taking the IC from E5, and performing a forward simulation.
The original operating condition is reliably recovered from both targets across a range of initial $T(t)$ guesses (\Cref{exFig3}(a)).

We then target a real experimental trajectory using the held-out experiment E5.
With the initial condition fixed at its measured value, the optimised temperature profile departs from the experimental one (MAE = 0.11 \unit{\degreeCelsius}) and the length and width trajectories are not accurately captured (\Cref{exFig3}(b)).
This mismatch is most noticeable in the length trajectory where the experimental final average length is not reached.
This is the same failure mode seen in \Cref{ssec-model_building} (treating the IC as ground truth), now at inference rather than training: without assimilation, $T(t)$ deviates to compensate for the IC mismatch (dashed blue line, \Cref{exFig3}(b)).
Freeing the initial condition for assimilation recovers a $T(t)$ profile that closely matches the experimental one (MAE = 0.06 \unit{\degreeCelsius}) and the target $\overline{L}_1$, $\overline{L}_2$ trajectories (\Cref{fig3}(a)).
Since there is no growth model here, the $T(t)$ profile does not match below $T_\mathrm{sat}$, which is the highlighted grey region in \Cref{fig3}(a).
The fact that the optimised profile is similar to the experimental profile for an unseen experiment supports that the model has learned to represent the underlying dynamics rather than fit noise in the training data.

Conventional workflows are not suited to this task: they treat the measured IC as exact, which biases the recovered profile, and with numerical gradients, assimilating it alongside $T(t)$ would require $\approx 4.4\times10^4$ forward solves per iteration.
With AD in our framework, the full optimisation takes less than \qty{5}{\min}.

\subsubsection{Product Aspect Ratio Optimisation} \label{sssec-aspect-ratio-optimisation}

Next, we shift the focus to an unknown answer and optimise towards a specific process product specification using the combined growth-dissolution model (\Cref{ssec-extended_kinetic_model}) - a forward design problem.
The product specification we aim to improve is the particle aspect ratio given by the ratio of the average crystal length and widths, $\overline{L}_1/\overline{L}_2$.
Downstream packing, drying, filtration \cite{periniFilterabilityPredictionNeedlelike2019}, and bioavailability of the solid product obtained through crystallisation \cite{lovetteCrystalShapeEngineering2008a, kocevskaChangingParadigmIndustrial2024} are affected by this.
Conventionally these profiles would be restricted to a few predefined low-dimensional shapes, such as giving the number and amplitude of temperature cycles.
In contrast, our framework takes the product specification directly as the objective and optimises end-to-end through the PBE to the \textit{full} temperature profile, so the search is driven by the product requirement rather than by the choice of profile family.

Starting from the initial condition of E5, we minimise the final product aspect ratio subject to the constraint that the product volume does not fall below the seed volume, $V_\mathrm{product}/V_\mathrm{seed}\geq1$, so that no material is lost.
Often temperature cycling, alternating between growth and dissolution regimes, is used for this kind of problem \cite{snyderManipulationCrystalShape2007} so we initialise our operating condition as a linear ramp that starts in dissolution and ends in growth.

\begin{figure}[h]
    \centering
    \includegraphics[width=\linewidth]{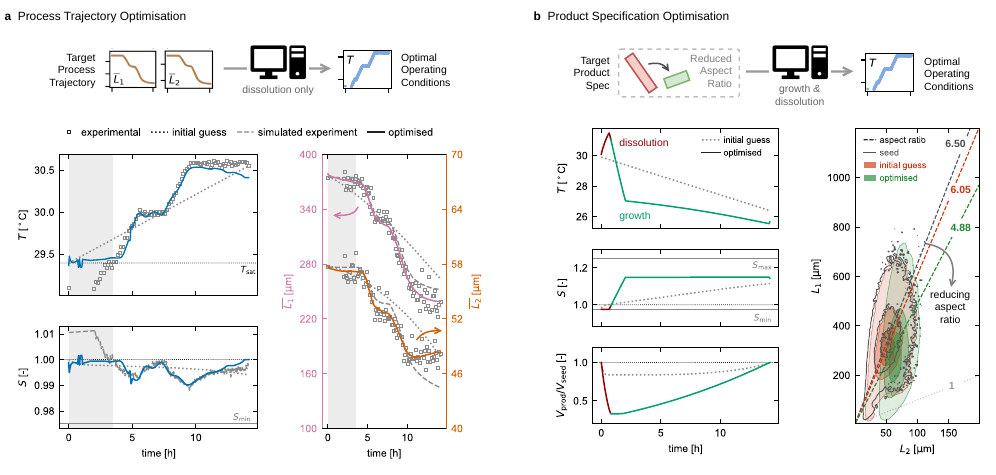}
    \caption{
        \textbf{End-to-end optimisation of operating conditions recovers target process trajectories and proposes new conditions to meet a product specification.}
        \textbf{(a)} Process trajectory optimisation recovers the operating profile of an unseen experiment.
        Using the frozen hybrid dissolution model (\Cref{fig2}), $T(t)$ and the initial condition ($f_0, c_0$) are optimised (solid blue) from an initial guess (dotted grey) to match the observed $\overline{L}_1$ (pink) and $\overline{L}_2$ (orange) trajectories of the held-out experiment E5 (open squares). 
        The recovered $T(t)$ closely tracks the experimental profile in the dissolution region despite never being an optimisation target (MAE = \qty{0.06}{\degreeCelsius}, vs \qty{0.11}{\degreeCelsius} with the measured IC, \Cref{exFig3}(b)), and $S(t)$ remains above $S_{\min}$ (0.975) throughout.
        Shaded grey: experimental temperature below $T_\mathrm{sat}$, which cannot be recovered as the hybrid model has no growth term.
        \textbf{(b)} Product specification optimisation reduces aspect ratio without net mass loss.
        Using the size-dependent growth \& dissolution model from the initial condition of E5, $T(t)$ is optimised (solid) from a linear dissolution $\to$ growth initial guess (dotted) to minimise the product aspect ratio, subject to $S_{\min}$, $S_{\max}$ and $V_\mathrm{product}/V_\mathrm{seed} \geq 1$.
        The optimised profile first dissolves the seed ($S<1$, red; volume falls to $\approx$35\%) before switching to growth ($S>1$, green; volume recovers to 100\%).
        Seed (grey line), optimised product (green contour) and initial-guess product (red contour) PSSDs are shown, with dotted lines marking each aspect ratio ($\overline{L}_1/\overline{L}_2$): 6.50, 4.88 (25\% reduction) and 6.05, respectively.
        Simulated result; experimental validation is beyond the scope of this work.
    }
    \label{fig3}
\end{figure}

This optimisation, shown in \Cref{fig3}(b), yields a design of $T(t)$ for this system that reduces the product aspect ratio from \num{6.50} to \num{4.88}, an improvement of 25\% from seed to product.
The initial guess profile gave a ratio of \num{6.05}.
The optimised profile in this case retains the single cycle, first dissolving at constant supersaturation, with the product volume reducing to around 35\%, before entering a growth region where the volume increases to 100\%.
This stays within the supersaturation limits and satisfies the product volume specification, both shown in \Cref{fig3}(b).
Because length growth scales with $L_1^{1.2}$ while width growth is size-independent, initially dissolving the population reduces subsequent length growth relative to width growth, redistributing mass towards the width as the volume is recovered.
It must be noted that this result is a purely model-generated hypothesis that requires experimental validation (beyond the scope of this work).

\section{Discussion} \label{sec-discussion}

This work demonstrates how end-to-end differentiable modelling extends mechanistic PBE solvers beyond forward simulation to a broader class of modelling, inference, and optimisation tasks.
The same solver and governing equations support forward simulation, recovery of a physical law, assimilation of the initial conditions that law is learned from, and optimisation of the operating conditions that drive the process.
Only the free variables $\boldsymbol{\upphi}$ change between these tasks, whether to diagnose the data ($\boldsymbol{\upphi} = [f_0, c_0]$), to build a model ($\boldsymbol{\upphi} = \boldsymbol{\uptheta}$), or to design an experiment ($\boldsymbol{\upphi} = T(t)$).
Differentiability extends the role of mechanistic solvers beyond forward simulation and simple parameter estimation, enabling data diagnosis, high-dimensional optimisation, and physical consistency checks.

A significant finding is that improved predictive performance is not always enough to recover a transferable, generalisable physical law.
When uncertain experimental data are treated as ground truth, a flexible network coupled to a PDE that is sensitive to its initial conditions cannot distinguish errors in the data from the underlying dynamics, and fits both.
Notably, this failure is only visible because the learned component is a small, inspectable function within a mechanistic model; a black-box model of the whole system would abstract it away.
This type of behaviour can therefore be hidden when ML is evaluated only on predictive performance, particularly in crystallisation and the wider chemical separations community, where ML methods are being increasingly adopted \cite{maRecentAdvancesDigital2026a, limaApplicationsMachineLearning2025, xiourasApplicationsArtificialIntelligence2022, chewAIChemicalEngineering2026, aprileReflectingBarriersContinuous2025}.

Treating the initial conditions as free variables separates error in the rate law from error in the data to which it is fitted, allowing the source of inconsistency to be diagnosed without additional experiments.
This matters beyond the present dataset: initial conditions, particularly for the crystal phase, are noisy and often simplified measurements from instruments that frequently disagree with one another and with ground truth \cite{biriThereRightParticle2025}, yet they are routinely taken as exact.
The resulting learned law was obtained in a single training effort and departs from the empirical model where supersaturation is densely sampled, suggesting that the observed non-linearity is present in the data rather than simply imposed by the NN.

Because the model is differentiable end-to-end, we can extend the use of the model inference time for operating modes that were not part of training, such as recovering the operating profile of an unseen experiment.
This differs from  black-box and surrogate approaches discussed in \Cref{sec-intro}, which are generally tied to the tasks and quantities that they were trained on.
The same approach applies to a product specification, where the framework returns both a candidate operating profile and an interpretable mechanism for achieving it.
This replaces the manual specification of operating conditions with gradient-based design, mirroring the way we replace selection of empirical rate laws with flexible model fitting.

The aspect ratio result is limited to a simulated hypothesis.
More generally, the demonstration is limited to a single material system and thirteen experiments, and the learned rate is size-independent.
Nevertheless, there is no additional data requirement compared with empirical modelling, the NNs remain small and straightforward to train, and all tasks required a fraction of the computational cost of conventional approaches.
Embedding the NN within the PBE solver also ensures that the governing conservation laws are satisfied.
The same framework can in principle be extended to size-dependent kinetics, to other PBE mechanisms (nucleation, breakage, agglomeration), and to other transport equations.
It is this generality, rather than any single result, that is the key contribution of this work.

\section{Methods} \label{sec-methods}

In this section we describe in detail our hybrid modelling framework.
All results are obtained using a single two-dimensional finite-volume PBE solver coupled to a solute mass balance equation to numerically model process trajectories (\Cref{ssec-modelling_methods}). 
Across tasks, only the convection term varies in the mathematical model, taking either the published empirical kinetic laws (\Cref{ssec-empirical_kinetic_models}) or a neural-network representation (\Cref{ssec-nn_methods}). 
When training the model, the training loss is minimised by gradient descent through the solver, with tasks only differing in their learnable (free) variables $\boldsymbol{\upphi}$: rate law parameters and initial conditions for model building (\Cref{ssec-training_methods}), and the temperature profile for process optimisation (\Cref{ssec-process_optimisation_methods}). 
The experimental dataset and train–validation split are described in \Cref{ssec-experimental_data_methods}, with computational benchmarking described in \Cref{ssec-computational_benchmarking}.

\subsection{Modelling} \label{ssec-modelling_methods}
We model a batch crystallisation system where there is no flow in or out of the crystallizer during each experiment.
The PBE that we use to simulate this system is a partial differential equation given by
\begin{equation} \label{eq-general_pbe}
		\frac{\partial f(\mathbf{L}, t)}{\partial t}
		+ \nabla_{\mathbf{L}} \cdot [U(\mathbf{L}, \mathbf{Y}, t)f(\mathbf{L}, t)]
		= 0
\end{equation}
where $f(\textbf{L},t)$ is the number density of crystals with characteristic lengths $\mathbf{L} = [L_1, L_2]$ at time $t$, and $\mathbf{Y}$ is the system state, capturing information such as the temperature and the concentration.
The rate at which crystals grow or dissolve (the convection rate) is given by $U(\textbf{L}, \mathbf{Y}, t)$.

The PBE describes the solid phase (i.e., the crystal population) and must therefore be coupled with the following mass balance to describe the liquid phase:
\begin{equation} \label{eq:general mass balance}
        \frac{\mathrm{d} c}{\mathrm{d} t}
        = -\rho_{\mathrm{c}} k_{\mathrm{v}} \frac{\mathrm{d} V_{\mathrm{c}}}{\mathrm{d} t}
\end{equation}
where $c$ [\unit{\gram \per \kg}] is the solute concentration in the liquid phase, $\rho_{\mathrm{c}} = 1.59 \times 10^{-12}$ [\unit{\gram \per \micro\meter\cubed}] is the crystal density, $k_{\mathrm{v}} = \pi/4$ [-] is the shape factor accounting for non-cuboidal crystal shapes, and $V_{\mathrm{c}}$ [\unit{\micro\meter\cubed \per \kg}] is the total crystal volume calculated by integrating the distribution.

These coupled equations are subject to the following initial and boundary conditions
\begin{align} \label{eq:ic and bc}
		f(\mathbf{L}, t=0) &= f_0 \\
		c(t=0) &= c_0 \\
		f(\mathbf{L} = 0, t) &= 0 \\
		f(\mathbf{L} = \infty, t) &= 0
\end{align}
where $f_0$ is the initial distribution, commonly referred to as the seed distribution, and $c_0$ is the initial concentration.

\subsection{Differentiable, GPU-Accelerated Numerical Solver}

We solve these equations using a fully discretised  finite volume method (FVM) \cite{levequeFiniteVolumeMethods2002}  for the internal (size) coordinates and forward Euler with a CFL condition for the temporal coordinates.
We build upon a previous solver written in JAX \cite{alsubeihiModernEfficientDifferentiable2025}.
Importantly, this solver is end-to-end differentiable, which unlocks our ability to do the tasks shown in this paper, namely NN optimisation, IC optimisation, and end-to-end process optimisation.

Automatic differentiation is the key feature of a differentiable solver and is used to replace numerical differentiation.
For numerical derivatives, forward differences for example, a forward pass needs to be solved for every individual gradient involved in an optimisation.
While this might be practical for small parameter numbers, the computational cost increases rapidly as the number of parameters increases, particularly if the underlying solver is not computationally efficient.
This becomes particularly important for the more complex optimisation tasks considered in this work, where many parameters may need to be optimised simultaneously.
In contrast, automatic differentiation propagates derivatives directly through the computational graph of the solver, avoiding the need for a separate perturbed forward simulation for each parameter.

Furthermore, the solver is efficiently implemented for GPUs, and has already been shown to outperform MATLAB, \texttt{numpy} and \texttt{C++} \cite{alsubeihiModernEfficientDifferentiable2025}.
We extend this solver by including NN components, allowing data assimilation, and utilising \texttt{vmap} to simulate many experiments at once, enabling training our model across the entire dataset simultaneously.
Further details on the solver implementation can be found in our previous work \cite{alsubeihiModernEfficientDifferentiable2025} which details the JAX solver.

\subsection{Empirical Kinetic Models} \label{ssec-empirical_kinetic_models}
The convection term $U(\mathbf{L}, \mathbf{Y}, t)$ is the system specific part of the model.
We consider models that are functions of temperature $T$ [\unit{\degreeCelsius}] and supersaturation $S$ [-], given at a time $t$ by
\begin{equation} \label{eq:supersaturation}
    S = \frac{c}{c^*},
\end{equation}
where $c^*$ [\unit{\gram \per \kilogram}] is the system specific solubility as a function of temperature given by 
\begin{equation}
    c^*(T)=3.37 e^{0.036T}
\end{equation}

B\"otschi et al.~\cite{botschiFeedbackControlSize2019} proposed and fitted four different empirical dissolution rate expressions to their experimental data, and we use two of these retaining their original names.
These are the two parameter size-independent linear model M1, which serves as our predictive performance benchmark, and the four parameter size-dependent non-linear model M4.
Both are used at their published parameters alongside a size-dependent growth model from Ochsenbein et al.~\cite{ochsenbeinGrowthRateEstimation2014} which is combined with M4 and used in \Cref{ssec-process_optimisation}.
This growth model was fitted for the same system of \textsc{l}-glutamic acid in water.

Each of these models is only used over the range in which it was originally fitted. 
These ranges define the supersaturation limits applied during the optimisations in \Cref{ssec-process_optimisation}. 
For M1 and M4, this is $S \geq S_{\min} = 0.975$ \cite{botschiFeedbackControlSize2019}, corresponding to the range of supersaturations observed in the experimental data. 
For the size-dependent growth model, this is $S \leq S_{\max} = 1.25$, as reported in their work \cite{ochsenbeinGrowthRateEstimation2014}.

\subsubsection{M1 - Predictive Performance Baseline}
The linear M1 model was the chosen model of B\"otschi et al.~\cite{botschiFeedbackControlSize2019} and acts as our predictive performance baseline in \Cref{ssec-model_building}.
This is also used in \Cref{exFig4} for computational benchmarking.
The M1 model describes the rates in each dimension, $D_1$ and $D_2$, as a function of supersaturation, $S$:
\begin{align} \label{eq:M1_coefficients}
		D_1 &= -2.32 (1-S) \\
		D_2 &= -0.19 (1-S)
\end{align}

\subsubsection{Size-Dependent Growth and Dissolution Model} \label{ssec-extended_kinetic_model}
We use an extended combined growth-dissolution kinetic model for the aspect ratio optimisation task in \Cref{ssec-process_optimisation}.
In this case two models are combined where the dissolution equation is used for $S<1$, and the growth equation is used for $S>1$.

Dissolution is the M4 model from B\"otschi et al. \cite{botschiFeedbackControlSize2019} given by
\begin{align} \label{eq:M4_coefficients}
    D_1 &= - 2.24 (1 - S) \left( 1 + L_1 \right) ^{-5.05\times10^{-2}} \\
    D_2 &= - 6.43 \times 10^{-4} (1 - S) \left( 1 +L_2 \right) ^{1.29}
\end{align}
This is the four parameter model shown in \Cref{exFig4}.

The growth model is from \cite{ochsenbeinGrowthRateEstimation2014} and is size-dependent in just the $G_1$ growth rate.
This model uses temperature in K and is given by
\begin{align} \label{eq:growth_model_coefficients}
    G_1 &= \exp \left(-\frac{6.9 \times 10^4}{T^2 \ln(S)}\right) 
    \exp \left(-\frac{1.9 \times 10^3}{T}\right) 
    (S-1)^{2/3} (\ln(S))^{1/6} L_1^{1.2} \\
    G_2 &= (S-1)^{2.7} \exp \left( - \frac{7 \times 10^2}{T} \right)
\end{align}

\subsection{Hybrid Dissolution Rate} \label{ssec-nn_methods}
The hybrid dissolution rate is parametrised as a multilayer perceptron (MLP), mapping two state variables, $S$ and $T$, to two dissolution rates, $\textbf{D} = [D_1, D_2]$:
\begin{equation} \label{eq:pure_nn}
		\textbf{D} = \mathrm{NN}_{\boldsymbol{\uptheta}}(S, T)
\end{equation}
where $\boldsymbol{\uptheta}$ denotes the weights and biases of the network.
The network has 2 input neurons, 2 output neurons, \texttt{tanh} activated hidden layers, and is clipped to return zero for cases where $S\geq1$.
This enforces no dissolution under supersaturated conditions.
A 3×20 architecture was selected from a preliminary sweep and smaller networks (1×5, 2×5) give similar accuracy (\Cref{exFig2}).

Each input $S$ and $T$ value is standardised before being passed to the network, mapping a raw data value $x$ to 
\begin{equation} \label{eq:standardization}
		x^* = \frac{x - \mu_x}{\sigma_x}
\end{equation}
where $\mu_x$ and $\sigma_x$ are the mean and standard deviation of all observed values of $x$ in the experimental data set.

The output layer uses a negative \texttt{softplus} activation to ensure that the rates predicted are negative.
Whilst the clip enforces zero dissolution for $S\geq1$, no structural element of the network forces the rate to approach zero as $S\to1$.
Since dissolution rates are not directly measurable, this is scaled by an estimated reference rate.
The reference rates are estimated as the mean change in average particle length and width per unit time across all experiments, e.g. $\dot{L}_{1,\mathrm{ref}} = \Delta \bar{L}_1 / \Delta t$.

\subsection{Experimental Data \& Split} \label{ssec-experimental_data_methods}

The experimental procedure used to generate the 13 experiments (E1 to E13) and data used was obtained by the senior author of this work.
This is detailed by B\"otschi et al. \cite{botschiFeedbackControlSize2019, botschiAlternativeApproachEstimate2018}.
A key enabler for this data was the DISCO \cite{rajagopalanComprehensiveShapeAnalysis2017}, which facilitates monitoring the size and shape of crystals in real time during the experiments, as shown in \Cref{fig1}(a).

The data split used was the same as that of B\"otschi et al. with two experiments from the validation set (E5, E12) left aside for the process optimisation task in \Cref{ssec-process_optimisation}.
For the standard training approach, this is the split that was used.
For the approach where the initial conditions were assimilated (\Cref{ssec-model_building}), the entire dataset was used for training (again, without E5 or E12), since each initial condition needed to be individually calibrated.

Key time series from the experimental data are shown in \Cref{exFig1} with each experiment labelled with its experiment number, seed mass, and whether it is part of the training, validation, or held out set.
The operating condition $T(t)$ and the three experimental targets ($c$, $L_1$, $L_2$) are shown.

\subsection{Hybrid Dissolution Rate and Initial Condition Training Procedure} \label{ssec-training_methods}

Three time resolved targets (process outputs) are used for training: the concentration of the molecules dissolving from the solid phase into the liquid phase ($c$), and the average particle length and width ($\overline{L}_1$, $\overline{L}_2$).
Examples of these targets from the experimental dataset are shown in \Cref{fig1}(a) and the full dataset is shown in \Cref{exFig1}.
These are used in the MSE loss function
\begin{equation} \label{eq:loss}
		\mathcal{L}(\boldsymbol{\upphi}) = \frac{1}{3N} \sum_{i=1}^{3} \sum_{j=1}^{N}
		(y_{i,j} - \hat{y}_{i,j}(\boldsymbol{\upphi}))^2,
\end{equation}
where each target is equally weighted, $y$ is an experimental data point, and $\hat{y}(\boldsymbol{\upphi})$ is a model prediction obtained from the FVM solver as a function of the free variables $\boldsymbol{\upphi}$.
Experimental data and predictions are standardised per target before calculating the loss using the same standardisation as \Cref{eq:standardization}, with $\mu$ and $\sigma$ calculated over the full dataset ahead of training.

We consider two training cases, both minimising $\mathcal{L}$ over the free variables $\boldsymbol{\upphi}$.
In \textit{Case A}, only the dissolution model parameters, $\boldsymbol{\upphi} = \boldsymbol{\uptheta}$, are optimised, while in \textit{Case B}, the initial conditions $f_0$ and $c_0$ are assimilated and included as additional free variables to account for experimental uncertainty in the measured initial conditions, $\boldsymbol{\upphi} = [\boldsymbol{\uptheta}, f_0, c_0]$:
\begin{align}
\text{Case A:} \quad & \min_{\boldsymbol{\uptheta}} \,\, \mathcal{L} \\
\text{Case B:} \quad & \min_{\boldsymbol{\uptheta}, f_0, c_0} \,\, \mathcal{L}
\end{align}
In both cases an Adam (gradient based) optimizer (\texttt{optax} \cite{deepmindDeepMindJAXEcosystem2020}) is used to minimise $\mathcal{L}$ with a fixed learning rate of $10^{-3}$ applied to $\boldsymbol{\uptheta}$, and in Case B this also applies to $f_0$ and $c_0$.
A \texttt{ReLU} is applied to $f_0$ at each update step to enforce non-negativity, since a particle size distribution is strictly non-negative.
This, combined with the fact that the initial distribution is rescaled to the correct mass (\Cref{exFig1}) during simulation, ensures that $f_0$ remains physical and stable.
For \textit{Case A}, all models are trained for up to 500 epochs, with the final model parameters taken from the epoch yielding the smallest validation loss.
Early stopping is implemented such that if the validation loss does not decrease over 100 epochs, training is terminated.
In \textit{Case B}, there is no data split so early stopping is based on the training loss rather than a separate validation loss.

\subsection{Process Optimisation}  \label{ssec-process_optimisation_methods}

\paragraph{Parametrisation}
As in \Cref{ssec-training_methods}, $\boldsymbol{\upphi}$ denotes the free variables optimised in a given task.
The key variable optimised here is the operating condition $T(t)$ that is defined by $300$ linearly spaced values in time across the length of each experiment.

\paragraph{Bounds and Penalties}
Two classes of constraint are applied: the operating conditions are restricted to what the equipment can deliver, and the supersaturation is restricted to the range over which the kinetic model in use is valid (\Cref{ssec-empirical_kinetic_models}).

The temperature is constrained to the range spanned by the experimental data, $20 - 41$\unit{\degreeCelsius}, as a hard constraint applied by box projection (\texttt{optax}) at every optimisation iteration. 
The remaining constraints are imposed as soft penalties in the loss function, $\mathcal{P}_{\dot T}$ and $\mathcal{P}_{S}$. 
$\mathcal{P}_{\dot T}$ limits the rate at which the temperature can change during an experiment and is set by the limits of the jacketed vessel: $|\mathrm{d}T/\mathrm{d}t| \leq 0.05$ \unit{\degreeCelsius \per \minute}. 
$\mathcal{P}_{S}$ keeps the supersaturation above $S_{\min}$ (0.975) throughout, and below $S_{\max}$ (1.25) wherever a growth law is present.
These are the limits of the data that each model was trained on.

\paragraph{Optimisation Procedure}
$\boldsymbol{\upphi}$ is optimised with Adam (\texttt{optax}, learning rate of $10^{-3}$) and is used to minimise the loss function
\begin{equation} \label{eq:generic-temp-loss}
	\mathcal{L}(\boldsymbol{\upphi}) = \mathcal{L}_{\mathrm{task}}(\boldsymbol{\upphi}) + \mathcal{P}_{\dot T} + \mathcal{P}_S,
\end{equation}
for 500 - 1000 iterations, where $\mathcal{L}_{\mathrm{task}}$ is defined per task, and $\mathcal{P}_{\dot T}, \mathcal{P}_S$ are the soft penalties.
In the two trajectory-tracking tasks, $\mathcal{L}_{\mathrm{task}}$ is the MSE between a target trajectory and the simulated one,
\begin{equation} \label{eq:opt-mse}
	\mathrm{MSE}\big(\overline{L}_1, \overline{L}_2\big) = \frac{1}{2N} \sum_{i=1}^{2} \sum_{j=1}^{N}
	\big(y_{i,j} - \hat{y}_{i,j}(\boldsymbol{\upphi})\big)^2,
\end{equation}
which takes the same form as the training loss (\Cref{eq:loss}) but over the two size targets only, with $y$ the target value at one of $N$ time points and $\hat{y}(\boldsymbol{\upphi})$ the prediction from the FVM solver.
The target is defined per task below.

\subsubsection{Synthetic Trajectory Optimisation} \label{sssec-synthetic-trajectory-optimisation-methods}
A target temperature profile, $T_{\mathrm{ref}}(t)$, is synthetically generated (as a stepped ramp or a smooth sigmoid) and simulated from the initial condition of E5 using the hybrid dissolution model (\Cref{ssec-model_building}), giving synthetic $L_1$ and $L_2$ targets.
The free variables are $\boldsymbol{\upphi} = T(t)$, and $\mathcal{L}_{\mathrm{task}}$ is the MSE in \Cref{eq:opt-mse} evaluated against the synthetic target trajectory.
This is repeated 20 times with unique randomised linear ramps, and the optimised profiles along with their spread and simulation results are shown in \Cref{exFig3}(a).

\subsubsection{Experimental Trajectory Optimisation} \label{sssec-experimental-trajectory-optimisation-methods}
Here, $\mathcal{L}_{\mathrm{task}}$ is the same MSE equation, \Cref{eq:opt-mse}, and is evaluated against smoothed measured $L_1$ and $L_2$ trajectories from E5 rather than synthetic targets.
Since the initial conditions of this experiment were not seen during training, $f_0$ and $c_0$ are included as free variables alongside $T(t)$: $\boldsymbol{\upphi} = [T(t), f_0, c_0]$.
This is also compared to a case where the initial condition is fixed in \Cref{exFig3}(b) and $\boldsymbol{\upphi} = T(t)$.
$T(t)$ is initialised as a ramp from the initial saturation temperature $T_\mathrm{sat}$ to the final recorded temperature since the hybrid model describes dissolution only.

\subsubsection{Product Aspect Ratio Optimisation} \label{sssec-aspect-ratio-optimisation-methods}
$T(t)$ is optimised from the initial condition of E5 using the combined growth-dissolution model from \Cref{ssec-extended_kinetic_model}.
$\boldsymbol{\upphi} = T(t)$, and the task is to minimise the product aspect ratio subject to a soft volume constraint:
\begin{equation}
\mathcal{L}_{\mathrm{task}} = \overline{L}_{1, \mathrm{product}}/\overline{L}_{2, \mathrm{product}} + \mathcal{P}_V,
\end{equation}
where $\mathcal{P}_V$ is a soft penalty enforcing $V_{\mathrm{product}}/V_{\mathrm{seed}} \geq 1$, with $V$ the total crystal volume, so that the product volume cannot fall below that of the seed.
$T(t)$ is initialised as a linear ramp descending from dissolution ($S<1$) into growth ($S>1$).

\subsection{Computational Benchmarking}  \label{ssec-computational_benchmarking}

We compare the computational performance of the original MATLAB solver from B\"otschi et al. \cite{botschiFeedbackControlSize2019} and our JAX based solver \cite{alsubeihiModernEfficientDifferentiable2025}.
Between the solvers we compare the M1 and M4 kinetic models, representing 2 and 4 optimised parameters.
We only report training times for M1 and M4 here for computational purposes - elsewhere in this work these models are used at their published coefficients \cite{botschiFeedbackControlSize2019}.
Additionally we show JAX results for the hybrid dissolution model from \Cref{ssec-model_building}, showing the case where we train only the NN and the case where we include the entire IC for assimilation.

\paragraph{Forward Simulation.}
We report the mean time required to perform a single forward simulation, averaged over the 13 experiments in the data set \cite{botschiFeedbackControlSize2019}.

\paragraph{Model Training.}
MATLAB uses the \texttt{fmincon} optimiser with numerically estimated gradients at default settings, run to its own convergence criterion. 
JAX uses the Adam optimiser (\texttt{optax}) with automatic differentiation with early stopping criteria (\Cref{ssec-training_methods}).
For the JAX models we report both (1) wall-clock time to convergence under early stopping, and (2) wall-clock time for a fixed budget of 500 epochs.
This isolates the convergence from how the computational cost scales with the number of parameters that are optimised and is not applicable to MATLAB which does not train in epochs.
The parameter counts, and all times are reported in \Cref{exFig4}.

\paragraph{Hardware.}
Both the JAX solver and MATLAB solver are run on the same machine for computational performance evaluation.
This computer is equipped with an AMD Ryzen 3900 12-core CPU and an NVIDIA RTX 4090 GPU.
The JAX solver was able to use the GPU automatically, whereas the MATLAB code was \texttt{.mex} compiled and does not support GPU execution.

\newpage
\bibliography{dissolution_hybrids_paper.bib} % common bib file

\newpage
\section*{Extended Figures}

\setcounter{figure}{0}
\renewcommand{\thefigure}{E\arabic{figure}}

\begin{figure}[h]
    \centering
    \includegraphics[width=\linewidth]{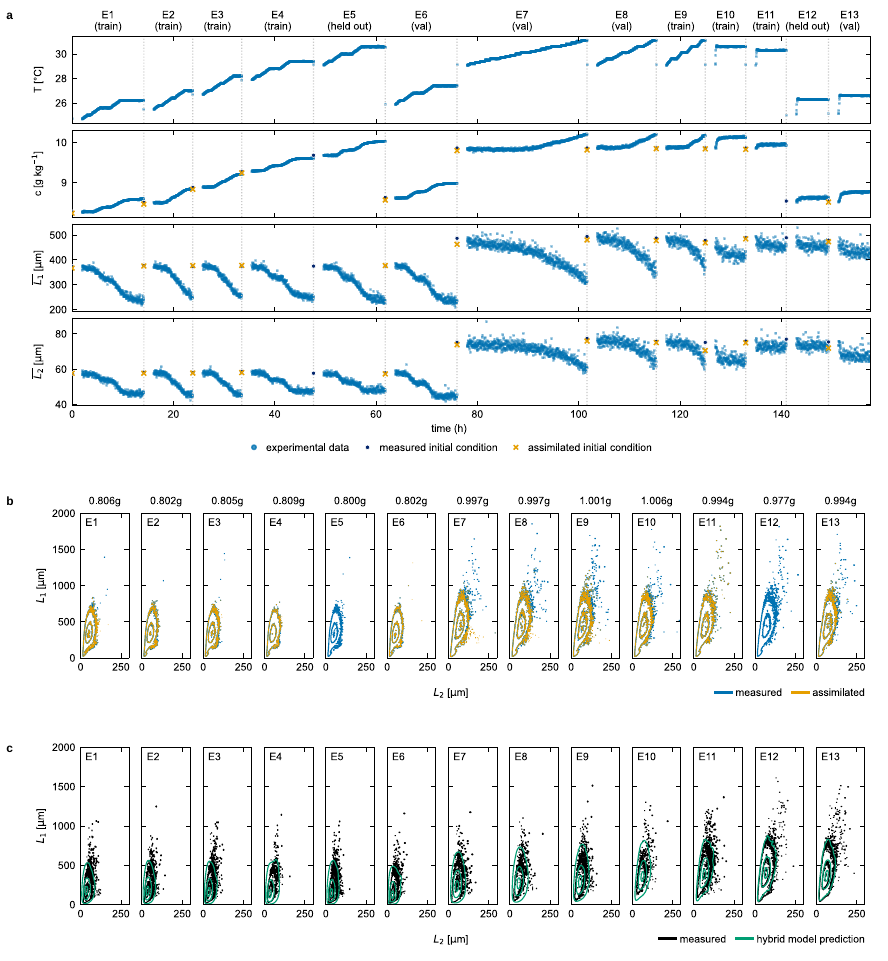}
    \caption{
            \textbf{Experimental dataset and initial condition assimilation.}
            (a) All experimental trajectories in the dataset (E1--E13), labelled by training/validation/held-out split, for the operating condition $T(t)$ and the three target metrics: solute concentration $c$, average crystal length $\overline{L}_1$, and width $\overline{L}_2$.
            For each target trajectory the measured initial condition (dark blue) and the assimilated initial condition (orange cross) are marked.
            (b) Full initial and assimilated particle size and shape distributions (PSSDs) for each experiment, with the corresponding experimental seed mass.
            Assimilated PSSDs are similar in shape to the original measurement.
            The two held out experiments do not have assimilated PSSDs, since they were not included in the hybrid model training.
            (c) Full final PSSDs from each experiment (black) along with the corresponding predicted PSSDs when simulating all of the experiments from the measured initial conditions using the hybrid dissolution model (from \textit{Case B}).
        }
    \label{exFig1}
\end{figure}

\newpage
\begin{figure}[h]
    \centering
    \includegraphics[width=\linewidth]{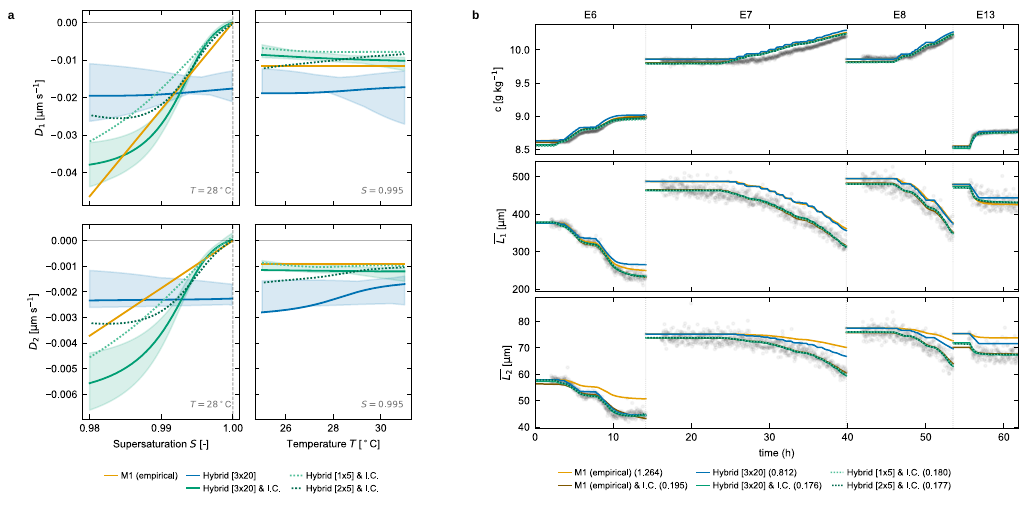}
    \caption{
            \textbf{Learned functional forms and predictive performance}.
            (a) Learned dissolution rate functional forms $D_1$ and $D_2$ vs.\ supersaturation $S$ (fixed $T=28\,^\circ$C, left) and vs.\ temperature $T$ (fixed $S=0.995$, right), for the M1 empirical baseline and hybrid models of varying architecture ($1\times5$, $2\times5$, $3\times20$ hidden layers $\times$ nodes), with and without joint IC assimilation. 
            Shaded bands give the spread across the 20-member ensembles where multiple models were trained.
            Importantly, this shows how the hybrid model trained on the measured initial condition can learn a non-physical form, with the rate never approaching zero at $S=1$.
            (b) Predicted vs.\ measured trajectories for the four validation experiments (E6, E7, E8, E13) across all three target metrics ($c$, $\overline{L}_1$, $\overline{L}_2$). 
            Legend values give the combined trajectory loss across these experiments for each model. 
            IC assimilation improves the fit for both the empirical (M1) and hybrid models, with the best performance coming from the hybrid model. 
            This is consistent across the different network architectures tested.
        }
    \label{exFig2}
\end{figure}

\newpage
\begin{figure}[h]
    \centering
    \includegraphics[width=\linewidth]{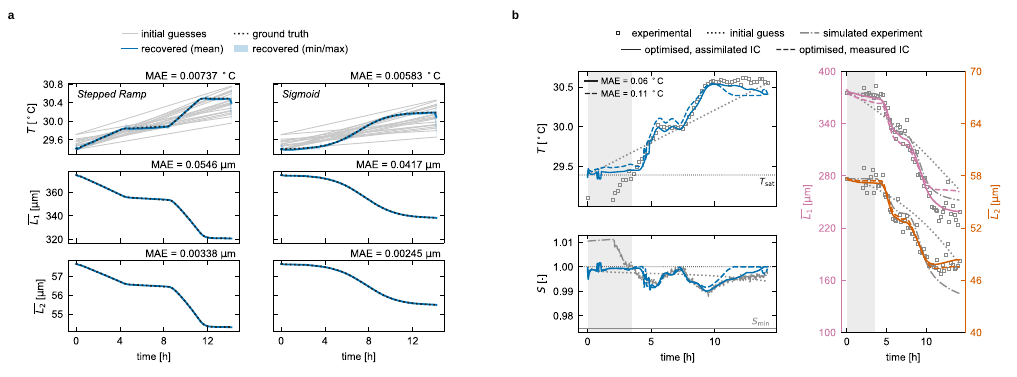}
        \caption{
            \textbf{Optimisation verification.}
            (a) Synthetic recovery of the operating condition $T(t)$ for two target profiles (stepped ramp, sigmoid), using the frozen hybrid dissolution model with the IC fixed at its measured value (E5). 
            Grey lines show 20 random linear-ramp initial guesses.
            The recovered mean (blue) and min/max envelope (shaded) closely match the known ground truth (dotted) in $T(t)$ and in the resulting $\overline{L}_1$, $\overline{L}_2$ trajectories, with MAE as reported. 
            This isolates the optimiser's ability to recover a known solution from any error due to model or IC uncertainty.
            (b) Experimental trajectory optimisation on the held-out experiment E5.
            Top Left: optimised $T(t)$ when assimilating the IC (solid blue line) and when using the measured IC (dashed blue line) compared to the experimental measurements (open squares) and the initial guess used for the optimisation (dotted grey line).
            Grey region shows the region where the experiment was below the saturation temperature, and the MAE of the optimised profiles to the experiments is reported.
            Bottom Left: simulated $S(t)$ with the two optimised $T(t)$ profiles compared to a simulation using the experimental measurements (dashed grey).
            Right: corresponding $\overline{L}_1$ (pink) and $\overline{L}_2$ (orange) trajectories. 
            Freeing the IC for assimilation tracks the experimental trajectories more closely than the fixed-IC optimisation and finds an optimised $T(t)$ closer to the experimental measurements.
            }
    \label{exFig3}
\end{figure}

\newpage
\begin{figure}[h]
    \centering
    \includegraphics[width=\linewidth]{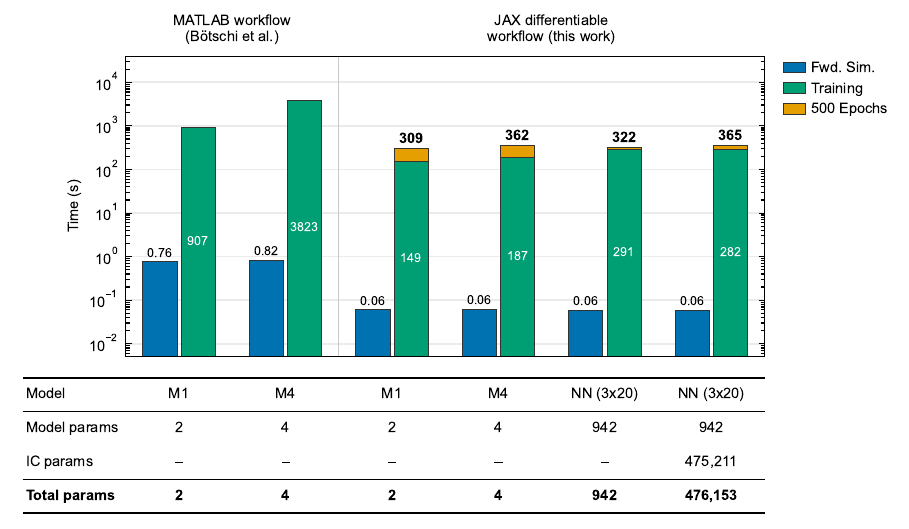}
        \caption{
            \textbf{Computational performance.}
            Wall-clock time (log scale) for forward simulation and optimisation, comparing the MATLAB workflow of B\"otschi et al. \cite{botschiFeedbackControlSize2019} against the JAX differentiable workflow of this work, for the M1 and M4 empirical models and the neural-network hybrid model ($3\times20$), with and without IC assimilation. 
            For JAX, the green segment gives time to convergence under early stopping, and the stacked orange segment (bold total above) gives the wall-clock time for a fixed 500 epochs.
            MATLAB does not train in epochs and has no equivalent segment. 
            The table gives the corresponding model, IC, and total parameter counts for each case. 
            Forward simulation is $\sim$13$\times$ faster in JAX (0.06\,s vs.\ 0.76 - 0.82\,s). 
            Adding $4.7\times10^5$ IC parameters for joint assimilation (NN $3\times20$, 942 $\to$ 476{,}153 total parameters) changes JAX convergence time only marginally (291\,s $\to$ 282\,s) and the fixed 500-epoch time by $<$15\% (322\,s $\to$ 365\,s), despite the five hundred fold increase in parameter count.
            This is consistent with gradient cost under automatic differentiation being effectively independent of parameter count. 
            MATLAB optimisation of the far smaller M1/M4 models takes 907\,s and 3823\,s respectively.
        }
    \label{exFig4}
\end{figure}
\newpage

\end{document}